\documentclass{PoS}

\title{Science Case of a Scintillator and Radio Surface Array at IceCube}

\ShortTitle{Science Case of the IceTop Upgrade}

\author{
The IceCube Collaboration\footnote{For collaboration list, see PoS(ICRC2019) 1177.}\\
{\itshape \href{http://icecube.wisc.edu/collaboration/authors/icrc19_icecube}{http://icecube.wisc.edu/collaboration/authors/icrc19\_icecube}}\\
E-mail: \email{fgs@udel.edu, frank.schroeder@kit.edu}
}

\abstract{
The upgrade of IceTop, the surface array of IceCube, by a hybrid array of scintillation and radio detectors is motivated by a rich science case.
The scintillators will lower the threshold for the measurement of air showers to about $100\,$ TeV, provide a more efficient veto of air showers for neutrino measurements, and improve the separation of the electromagnetic and muonic shower components due to the different responses of scintillators and ice-Cherenkov tanks. 
Furthermore, the scintillators will enable the calibration and compensation of the effect of snow accumulation above IceTop.  
The radio antennas will provide a calorimetric measurement of the electromagnetic shower component and a direct sensitivity to the shower maximum. 
Consequently, the combination of the existing ice-Cherenkov detectors in the ice and at the surface with the new scintillation and radio detectors at the surface will enable unprecedented accuracy for event-by-event mass classification in the PeV to EeV range.
This will transform IceCube into the most accurate instrument for high-energy Galactic cosmic rays in the Southern Hemisphere. 
Hence, the hybrid array will make an important contribution to the main science case of IceCube of understanding the origin of cosmic rays.
In addition to its cosmic-ray science goals, the hybrid array provides essential R\&D for IceCube-Gen2 which will feature a larger surface array, sophisticated timing and communication technology, and elevated surface structures. 
Moreover, the hybrid array will improve the understanding of the atmospheric background to the neutrino measurements suffering from uncertainties in the absolute flux and the mass composition, and from deficiencies of hadronic interaction models. 
Finally, the hybrid array opens new scientific opportunities, such as the searches for PeV gamma rays from the Galactic Center and for mass-dependent anisotropies, which both may lead to the discovery of the most energetic sources in the Milky Way.

\vspace{4mm}
{\bfseries Corresponding author:}
\speaker{Frank G. Schr\"oder}$^{,1,2}$\\
{$^{1}$ \itshape Bartol Research Institute, Dept.~of Phys.~and Astr., Univ.~of Delaware, Newark, DE, U.S.A.}\\
{$^{2}$ \itshape Institute of Nuclear Physics, Karlsruhe Institute of Technology (KIT), Karlsruhe, Germany}
}

\FullConference{36th International Cosmic Ray Conference -ICRC2019-\\
		July 24th - August 1st, 2019\\
		Madison, WI, U.S.A.}

\begin{document}

\section{Introduction}\label{sec:info}
Understanding the origin of cosmic rays is the primary science goal of the IceCube neutrino observatory at the South Pole. 
In particular, astrophysical neutrinos measured by the $1\,$km$^3$ large in-ice array \cite{IceCube:2018cha} can reveal sources of cosmic rays.
IceCube's astrophysical neutrinos seem to be dominated by extragalactic sources \cite{Albert:2018vxw}, and one extragalactic source has already been identified in a multi-messenger observation of neutrinos and gamma-rays \cite{IceCube:2018dnn}.
IceTop is the $1\,$km$^2$ surface array of IceCube above its in-ice detector \cite{IceCube:2012nn, Aartsen:2016nxy} and is comprised of 162 ice-Cherenkov detectors.
Icetop measures cosmic-ray air showers in the energy range of about $250\,$TeV to $1\,$EeV \cite{Aartsen:2013wda}, where the most energetic Galactic cosmic rays are presumed. 
Since both, the most energetic Galactic and extragalactic sources are unknown, IceTop and the in-ice detector complement each other in the quest for identifying the unknown cosmic-ray sources. 
Moreover, cosmic-ray measurements directly contribute to the neutrino science of IceCube by calibration of the in-ice detector with muons, the possibility to veto air showers, and by improving our understanding of the atmospheric background in the neutrino measurements.
Furthermore, IceTop addresses further science goals such as the search for PeV photons \cite{Aartsen:2012gka}, and the study of solar cosmic rays at low energies \cite{Abbasi:2008vr}.

IceTop has collected data for more than 10 years. 
Hence, many analyses are not limited by statistics, but by systematic uncertainties. 
In particular, the snow accumulation on the IceTop tanks increases the detection threshold for air showers and the systematic uncertainty because current models for the snow attenuation lack accuracy. 
Therefore, an enhancement of the surface detector by scintillation detectors was originally planned to mitigate the effect of the snow \cite{ScintillatorsICRC:2017}.
Recently, the geometry of the array and the station layout have been optimized to enable a significantly broader science case, and to reduce the deployment effort without deteriorating its performance.
By adding radio antennas to the same array, the accuracy and sky coverage for high-energy cosmic rays will be enhanced significantly beyond the original performance of IceTop.
Therefore, radio antennas were added to the existing prototype station at the South Pole \cite{IceScint_ICRC2019, IceRad_ICRC2019}.
In 2020, this prototype station will be refurbished to the new layout (see next section), and the installation of the complete array is planned for the subsequent years.\footnote{Moreover, there are plans to add small air-Cherenkov telescopes (IceAct) to further increase the accuracy for cosmic rays at lower energies \cite{IceAct_ICRC2019}.} 
While scintillators are an established technique for air-shower arrays, the radio technique has matured in the last years and has reached similar measurement accuracy as established optical techniques \cite{SchroederReview2016, HuegeReview2016}.
By this, the scintillator-radio upgrade of IceTop will enable new science goals as well as progress in the current goals of IceCube, such as key contributions to the science of the highest energy Galactic cosmic rays \cite{Astro2020_GCR_WhitePaper}.

\begin{figure}[t]
    \centering
    \includegraphics[width=0.99\linewidth]{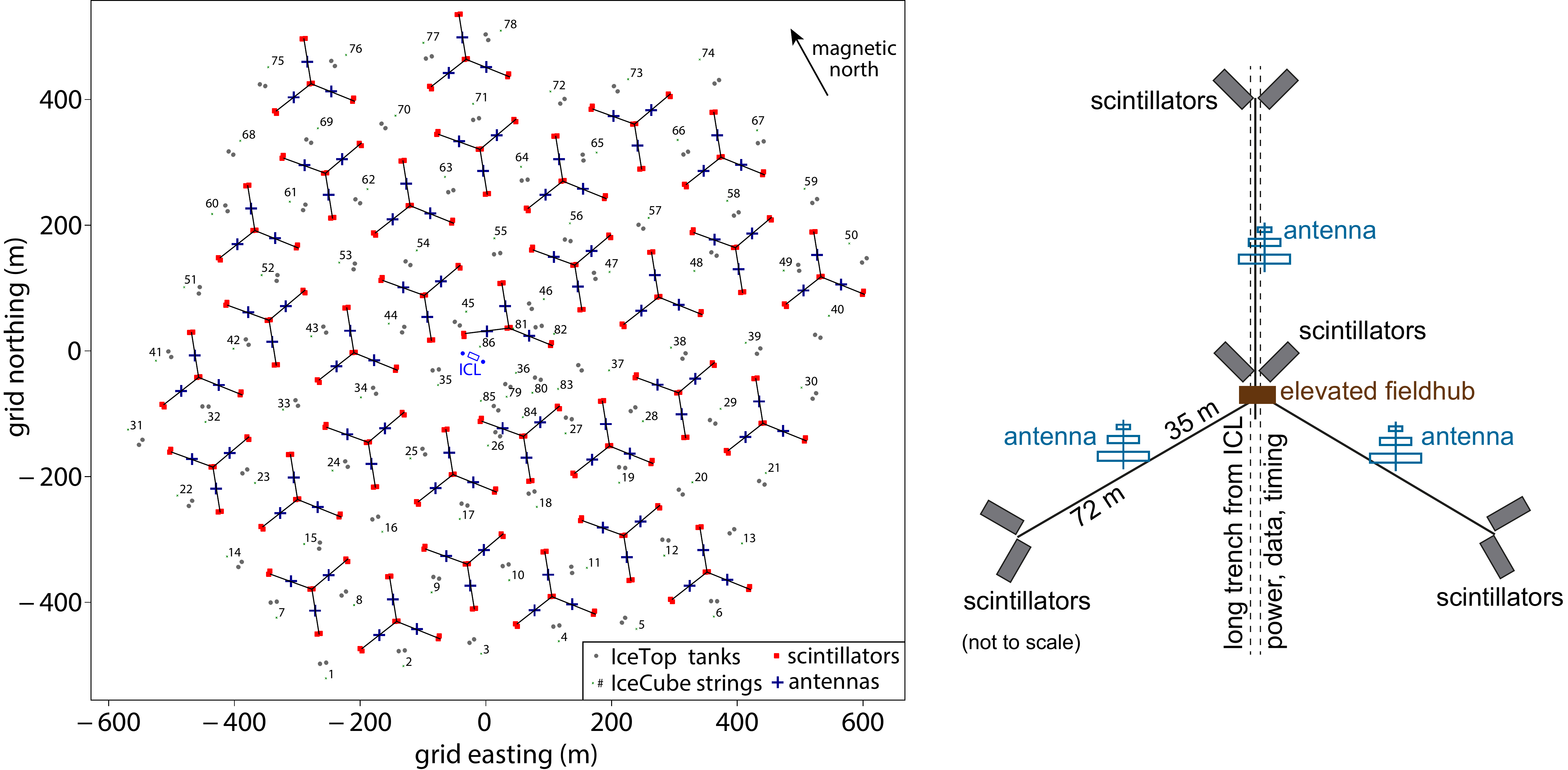}
    \caption{Conceptual layout of the scintillator-radio hybrid array (left) comprised of 32 stations (right). 
    Each station consists of 8 scintillation panels arranged in pairs, one pair at the center of the station where the local data-acquisition is located in an elevated field hub, and three pairs at $72\,$m distance from the station center. 
    Along the same spoke-trenches to these scintillators, three radio antennas with two polarization channels each will be deployed in $35\,$m distance to the center.
    }
    \label{fig:array}
\end{figure}

\section{Planned Detectors and Array Layout}
The scintillator-radio hybrid array will cover the existing IceTop array above the in-ice detector. 
It will consist of 32 stations comprised each of 4 pairs of scintillation detectors and 3 dual-polarized radio antennas connected to a local data-acquisition (DAQ) electronics at the center (Fig.~\ref{fig:array}). 
To avoid deterioration of the measurement response by snow and allow for easy maintenance access, all detectors, i.e. scintillator panels and antennas, as well as the fieldhub housing the central electronics will be elevated structures.
Each scintillation detector contains a SiPM whose signal is digitized locally upon a simple threshold trigger. 
Local triggers, digital signal amplitudes, and timestamps are then sent to the station DAQ. 
The radio signals are digitized directly in the station DAQ when receiving multiple coincident scintillation triggers.
Finally, the station DAQ transmits the data to the central DAQ in the IceCube Lab where data are stored on disk and transferred once yearly to the North. 
A limited amount of air-shower data, e.g., for veto or monitoring purposes, can be transferred additionally via satellite almost in real time.

\begin{figure}[t]
    \centering
    \includegraphics[width=0.99\linewidth]{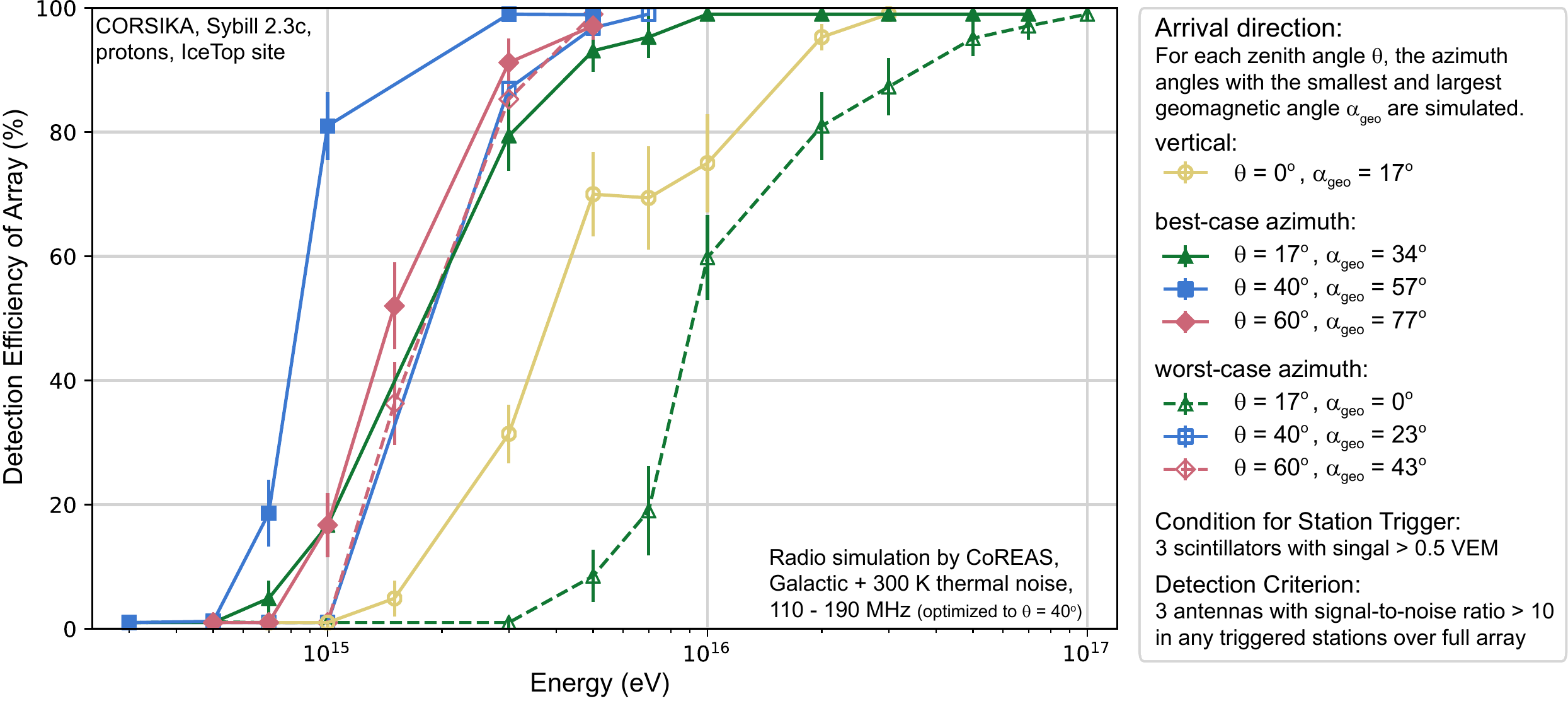}
    \caption{Detection threshold of the scintillator-radio hybrid array determined by CORSIKA proton simulations. 
    A signal-to-noise ratio > 10 is demanded in at least three antennas triggered by scintillators (noise model and normalization as in \cite{Balagopal2018}). 
    }
    \label{fig:radioThreshold}
\end{figure}

This is significantly lower than the threshold of IceTop, even before it was covered by snow.
The threshold of the radio array will be generally higher and depends on the zenith angle and the angle to the geomagnetic field. 
Simulation studies assuming Galactic and thermal noise of $300\,$K (noise model and normalization as in \cite{Balagopal2018}) predict an optimum frequency band including the FM broadcast band (fortunately this band is not in use at the South Pole) or at even higher frequencies.
The exact band can be chosen differently for each science case by digital filters during data analysis \cite{Balagopal2018}, e.g., the band of $110 - 190\,$MHz provides the lowest threshold for proton showers of $40^\circ$ zenith angle of around $1\,$PeV (see Fig. \ref{fig:radioThreshold}). 
For this study, a conservative radio threshold of a signal-to-noise ratio of $10$ is assumed motivated by other experiments \cite{Aab:2015vta, TunkaRex_NIM}, which will be lowered by applying more sophisticated techniques for pulse identification \cite{Erdmann:2019nie, Glaser:2019rxw, TunkaRexPRD2018}.
The future implementation of a radio self-trigger will be of particular advantage for the detection of inclined showers, e.g., from the direction of the Galactic Center which is continuously visible at $61^\circ$ zenith angle from the South Pole.

Thanks to the low radio background at the South Pole, a special feature of the radio array will be its sensitivity to showers parallel to the Earth's magnetic field, whose radio emission is an order of magnitude weaker and purely by the Askaryan effect.
This will enable to better study Askaryan emission, which is considered the only relevant emission mechanism for radio detection of neutrinos with in-ice arrays. 
It will also enable full efficiency and full sky coverage for the scintillator-radio hybrid array at the highest energies around $100\,$PeV and above.

\section{Science Goals}

The science case of the scintillator-radio enhancement of IceTop is rich and broad, pursuing several science goals regarding astrophysical neutrinos, cosmic rays and air showers, as well as multi-messenger astroparticle physics (see Table \ref{tab:my_label} and Refs.~\cite{Schroder:2018dvb, Haungs:2019ylq}). 
The most important goals are summarized in this section.

\subsection{More Accurate Air-Shower Measurements}
The surface enhancement will improve air-shower measurements in several ways by increasing the total accuracy and sky coverage and lowering the detection threshold compared to the present IceTop. 

First, the current IceTop suffers from snow accumulating above the tanks and attenuating air-shower particles, mostly the electromagnetic component. 
The limited understanding of the snow attenuation leads to significant systematic uncertainties in the interpretation of recent measurements.
This will be mitigated by a detailed calibration of the snow effect depending on snow height, zenith angle, and lateral distance. 
Moreover, the attenuation by snow approximately doubled the detection threshold of IceTop since its construction. 
By deploying the scintillators above the snow, the new detection threshold will be lowered by a factor of two compared to the original IceTop before snow coverage.

Second, hadronic interaction models can be tested much more precisely due to the separate measurements of the different shower components: particles at ground level measured by IceTop and the scintillators, high-energy muons measured by the in-ice detector, and the size of the electromagnetic component at shower maximum measured by the radio antennas. 
Since the signals of the particle detectors at the surface are in most cases dominated by electromagnetic particles, the different response to electrons and muons by the scintillators and tanks contains additional information on the muon number at ground level.
$X_\mathrm{max}$ measurements by radio may be used to identify proton-initiated showers \cite{Collaboration:2012wt}.
The ability to select proton events will enable stricter tests of hadronic models than possible today by studies based on the average mass composition \cite{Dembinski:2019uta}.

Third, the simultaneous measurement of several shower components by different detectors will boost the precision and accuracy.
Combining the classical methods of the electron-muon ratio and of $X_\mathrm{max}$ in one single experiment will yield unprecedented measurement accuracy for the mass composition and a per-event classification of the primary particle. 
Moreover, the combination of radio and muon measurements provides a new promising option for per-event mass classification \cite{Holt:2019fnj}.
Radio antennas can additionally be used to cross-check the absolute energy scale of IceCube with that of other air-shower arrays featuring radio antennas \cite{Apel:2016gws, Aab:2016eeq}.

These essential improvements in air-shower detection will enable a number of exciting science goals, in particular regarding Galactic cosmic rays and regarding IceCube's neutrino measurements.

\begin{table}[t]
    \centering
    \begin{tabular}{p{0.47\linewidth}|p{0.47\linewidth}}
        \textbf{Air Showers} & \textbf{Neutrino Science}\\
        \hline
        \vspace{-0.2cm} $\bullet$\,\,\,Mitigate and calibrate the snow attenuation of air showers measured by IceTop & \vspace{-0.2cm} $\bullet$\,\,\,More accurate estimation of atmospheric backgrounds for astrophysical neutrinos \\
        \vspace{-0.1cm} $\bullet$\,\,\,Better tests of hadronic interaction models & \vspace{-0.1cm} $\bullet$\,\,\,Improved surface veto for muon events \\
        \vspace{-0.1cm} $\bullet$\,\,\,Higher accuracy for mass and energy & \vspace{-0.1cm} $\bullet$\,\,\,Calibration of in-ice detector by muons\\
        &\\
        \textbf{Galactic Cosmic Rays} & \textbf{Technical Merits} \\
        \hline
        \vspace{-0.2cm} $\bullet$\,\,\,Search for PeV photons, in particular from H.E.S.S. source at Galactic Center \cite{Abramowski:2016mir, Balagopal2018} & \vspace{-0.2cm} $\bullet$\,\,\,Testbed for general infrastructure of IceCube-Gen2: communication, timing, etc. \\
        \vspace{-0.1cm} $\bullet$\,\,\,Understand transition to extragalactic CR& \vspace{-0.1cm} $\bullet$\,\,\,Development of elevated surface structures\\
        \vspace{-0.1cm} $\bullet$\,\,\,Mass composition measured more accurately + search for mass-sensitive anisotropy & \vspace{-0.1cm} $\bullet$\,\,\,Pathfinder for Gen2 surface array: in-ice calibration, veto aperture and threshold, etc.\\
    \end{tabular}
    \vspace{-0.1cm}
    \caption{Overview of the main scientific and technical goals of the IceTop upgrade (see text for details).}
    \label{tab:my_label}
\end{table}

\subsection{High-Energy Galactic Cosmic Rays}
\begin{figure}[t]
    \centering
    \includegraphics[width=13cm]{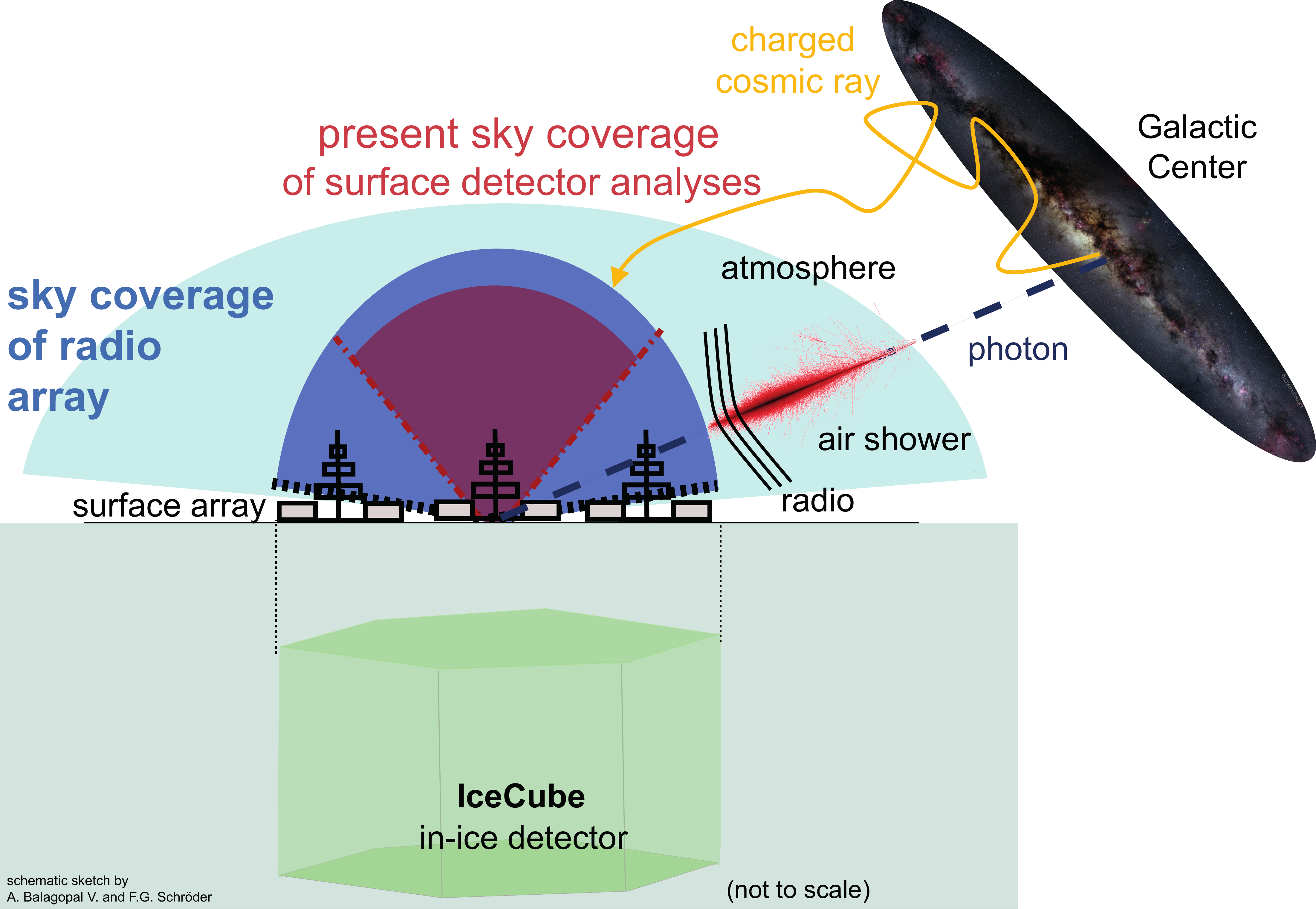}
    \caption{Schematic sketch of measurements with the IceTop upgrade: the scintillator enhancement will lower the energy threshold compared to IceTop and increase the precision. 
    The combination with radio antennas will provide a significant further improvement in total accuracy and lower the sky coverage bringing the Galactic Center in the field of view (figure from \cite{Schroder:2018dvb}).}
    \label{fig:radio_sketch}
\end{figure}

The surface enhancement provides an opportunity to solve the puzzle of the origin of the most energetic Galactic cosmic rays (see Fig.~\ref{fig:radio_sketch}). 
Depending on the scenario for the transition from Galactic to extragalactic cosmic rays, the highest energy particles still originating from our Galaxy may be at energies around $100\,$PeV up to a few EeV. 
This is supported by several observations, e.g., a hardening of the spectrum of light cosmic rays and a softening of heavy cosmic rays around $100\,$PeV \cite{KGheavyKnee2011, 2013ApelKG_LightAnkle}.
Since the maximum acceleration energy depends on the nuclear charge and mass numbers, the most energetic Galactic cosmic rays are presumed to be heavier nuclei, at least from the CNO group, probably from the Fe group. 
Thus, better measurements of the mass composition will help to determine the transition energy and mechanism. 

Anisotropy measurements may contain additional information on the number and distribution of Galactic sources contributing at the transition energy.
However, no significant anisotropy has been observed in this energy range, yet, only at lower and higher energies \cite{Aartsen:2018ppz, Apel:2019afz, Aab:2017tyv}.
This may be due to different phases of the large-scale Galactic and extragalactic anisotropies overlapping in the energy range of the transition.
Hence, the event-by-event mass classification provided by the upgraded surface array can be used to build a data set enriched by heavy nuclei, and thus presumably Galactic, cosmic rays of the highest energies.  
This will be used to extend the current anisotropy measurements by IceTop \cite{IceCube:2013mar} to higher energies and may lead to the first ever observation of mass-sensitive anisotropies. 

Finally, the enhanced accuracy for the primary particle as well as the improved sky coverage due to the radio array will improve the search for PeV photons.
In particular, the pevatron at the Galactic Center discovered by H.E.S.S. \cite{Abramowski:2016mir} is a promising candidate \cite{Balagopal2018}.
If the photon spectrum extends to higher energies, then the first detection of PeV photons may directly reveal a source, potentially \emph{the} source, of the most energetic Galactic cosmic rays.\footnote{The sensitivity to PeV photons from the Galactic Center will also open another channel for the search of heavy dark matter by IceCube \cite{Aartsen:2018mxl}.}

\subsection{High-Energy Neutrinos}
Atmospheric neutrinos produced in cosmic-ray air showers and interacting in the in-ice detector are a significant background to the measurement of astrophysical neutrinos. 
Therefore, this background needs to be understood as accurately as possible.
At the moment, significant uncertainties on the flux of atmospheric neutrinos \cite{Gaisser:2016obt}, each of several $10\,\%$, are caused by measurement uncertainties of the cosmic-ray flux versus the energy per nucleon and by deficiencies in the hadronic interaction models. 
Therefore, better tests of the hadronic models will help to solve the latter issue.
The measurement uncertainties on the absolute flux versus energy per nucleon will be reduced by the IceTop enhancement due to more accurate measurements of the mass composition and of the absolute energy scale. 
Consequently, the boosted accuracy for air-shower measurements will also improve the measurement of high-energy astrophysical neutrinos by improving the estimation of their background.

For the upper hemisphere of the sky, a more effective air-shower veto will enable the identification of additional astrophysical neutrino candidates interacting between IceTop and the in-ice detector \cite{IceTopVeto_ICRC2019}.
At the same time, muons are also an excellent calibration tool for the in-ice detector \cite{Bai:2007zzm}, which will benefit from more accurate measurements of the associated air showers. 

\subsection{Pathfinder for IceCube-Gen2}
The scintillator-radio array also serves as a pathfinder for a larger surface array planned for IceCube-Gen2, complementing the enlarged optical and radio in-ice arrays.
On the one hand, the IceTop Upgrade provides a testbed for technical solutions such as elevated surface structures, communication and timing infrastructure.
On the other hand, detector concepts will be tested, such as an efficient veto for near-vertical showers by scintillators, and for inclined showers by radio antennas.

\clearpage

\section{Conclusion}
In summary, there are significant synergies in enhancing the IceTop array by scintillators and radio antennas at the same time.
In the next year, the existing prototype station will be updated to the new design presented here, with the aim of deployment of the full array in subsequent years.
Together with the existing in-ice and surface detectors, this hybrid array will boost the accuracy of IceCube for air showers and enable a broad science case in the era of multi-messenger astronomy. 

\bibliographystyle{ICRC}
\bibliography{references}

\providecommand{\href}[2]{#2}\begingroup\raggedright\begin{thebibliography}{10}

\bibitem{IceCube:2018cha}
{\bf IceCube} Collaboration, M.~G. Aartsen et~al., {\em Science} {\bf 361}
  (2018) 147--151.

\bibitem{Albert:2018vxw}
{\bf ANTARES, IceCube} Collaboration, A.~Albert et~al., {\em Astrophys. J.}
  {\bf 868} (2018) L20.

\bibitem{IceCube:2018dnn}
{\bf IceCube, et al.} Collaboration, M.~G. Aartsen et~al., {\em Science} {\bf
  361} (2018) eaat1378.

\bibitem{IceCube:2012nn}
{\bf IceCube} Collaboration, R.~Abbasi et~al., {\em Nucl. Instrum. Meth.} {\bf
  A700} (2013) 188--220.

\bibitem{Aartsen:2016nxy}
{\bf IceCube} Collaboration, M.~G. Aartsen et~al., {\em JINST} {\bf 12} (2017)
  P03012.

\bibitem{Aartsen:2013wda}
{\bf IceCube} Collaboration, M.~G. Aartsen et~al., {\em Phys. Rev.} {\bf D88}
  (2013) 042004.

\bibitem{Aartsen:2012gka}
{\bf IceCube} Collaboration, M.~G. Aartsen et~al., {\em Phys. Rev.} {\bf D87}
  (2013) 062002.

\bibitem{Abbasi:2008vr}
{\bf IceCube} Collaboration, R.~Abbasi et~al., {\em Astrophys. J.} {\bf 689}
  (2008) L65--L68.

\bibitem{ScintillatorsICRC:2017}
{T. Huber for the \textbf{IceCube-Gen2} Collaboration},  \pos{PoS(ICRC2017)401}
  (2018).

\bibitem{IceScint_ICRC2019}
{\bf IceCube} Collaboration,  \pos{PoS(ICRC2019)309} (these proceedings).

\bibitem{IceRad_ICRC2019}
{\bf IceCube} Collaboration,  \pos{PoS(ICRC2019)401} (these proceedings).

\bibitem{IceAct_ICRC2019}
{\bf IceCube} Collaboration,  \pos{PoS(ICRC2019)179} (these proceedings).

\bibitem{SchroederReview2016}
F.~G. Schr\"oder, {\em Prog. Part. Nucl. Phys.} {\bf 93} (2017) 1--68.

\bibitem{HuegeReview2016}
T.~Huege, {\em Phys. Rept.} {\bf 620} (2016) 1--52.

\bibitem{Astro2020_GCR_WhitePaper}
F.~G. Schroeder et~al., {\em BAAS} {\bf 51} (2019) 131. Astro2020 Science White
  Paper.

\bibitem{Balagopal2018}
A.~{Balagopal V.} et~al., {\em EPJ C} {\bf 78} (2018) 111.

\bibitem{Aab:2015vta}
{\bf Pierre Auger} Collaboration, A.~Aab et~al., {\em Phys. Rev.} {\bf D93}
  (2016) 122005.

\bibitem{TunkaRex_NIM}
{\bf Tunka-Rex} Collaboration, P.~A. Bezyazeekov et~al., {\em Nucl. Instrum.
  Meth.} {\bf A802} (2015) 89--96.

\bibitem{Erdmann:2019nie}
M.~Erdmann, F.~Schl{\"u}ter, and R.~Smida, {\em JINST} {\bf 14} (2019) P04005.

\bibitem{Glaser:2019rxw}
C.~Glaser et~al., {\em Eur. Phys. J.} {\bf C79} (2019) 464.

\bibitem{TunkaRexPRD2018}
{\bf Tunka-Rex} Collaboration, P.~A. Bezyazeekov et~al., {\em Phys. Rev.} {\bf
  D97} (2018) 122004.

\bibitem{Schroder:2018dvb}
{F.G.~Schr\"oder for the \textbf{IceCube-Gen2} Collaboration}, {\it {Physics
  Potential of a Radio Surface Array at the South Pole (ARENA 2018)}},  in {\em
  {Acoustic and Radio EeV Neutrino Detection Activities (ARENA 2018) Catania,
  Italy, June 12-15, 2018}}, 2018.
\newblock \href{http://arxiv.org/abs/1811.00599}{{\tt arXiv:1811.00599}}.

\bibitem{Haungs:2019ylq}
{\bf IceCube} Collaboration, A.~Haungs, {\em EPJ Web Conf.} {\bf 210} (2019)
  06009.

\bibitem{Collaboration:2012wt}
{\bf Pierre Auger} Collaboration, P.~Abreu et~al., {\em Phys. Rev. Lett.} {\bf
  109} (2012) 062002.

\bibitem{Dembinski:2019uta}
{\bf EAS-MSU, IceCube, KASCADE-Grande, NEVOD-DECOR, Pierre Auger, SUGAR,
  Telescope Array, Yakutsk EAS Array} Collaboration, H.~P. Dembinski et~al.,
  {\em EPJ Web Conf.} {\bf 210} (2019) 02004.

\bibitem{Holt:2019fnj}
E.~M. Holt, F.~G. Schr{\"o}der, and A.~Haungs, {\em Eur. Phys. J.} {\bf C79}
  (2019) 371.

\bibitem{Apel:2016gws}
{\bf Tunka-Rex, LOPES} Collaboration, W.~D. Apel et~al., {\em Phys. Lett.} {\bf
  B763} (2016) 179--185.

\bibitem{Aab:2016eeq}
{\bf Pierre Auger} Collaboration, A.~Aab et~al., {\em Phys. Rev. Lett.} {\bf
  116} (2016) 241101.

\bibitem{Abramowski:2016mir}
{\bf H.E.S.S.} Collaboration, A.~Abramowski et~al., {\em Nature} {\bf 531}
  (2016) 476.

\bibitem{KGheavyKnee2011}
{\bf KASCADE-Grande} Collaboration, W.~D. Apel et~al., {\em Phys. Rev. Lett.}
  {\bf 107} (2011) 171104.

\bibitem{2013ApelKG_LightAnkle}
{\bf KASCADE-Grande} Collaboration, W.~D. Apel et~al., {\em Phys. Rev.} {\bf
  D87} (2013) 081101.

\bibitem{Aartsen:2018ppz}
{\bf HAWC, IceCube} Collaboration, A.~U. Abeysekara et~al., {\em Astrophys. J.}
  (2018). [Astrophys. J.871,96(2019)].

\bibitem{Apel:2019afz}
{\bf KASCADE-Grande} Collaboration, W.~{Apel} et~al., {\em Astrophys. J.} {\bf
  870} (2019) 91.

\bibitem{Aab:2017tyv}
{\bf Pierre Auger} Collaboration, A.~Aab et~al., {\em Science} {\bf 357} (2017)
  1266--1270.

\bibitem{IceCube:2013mar}
{\bf IceCube} Collaboration, M.~G. Aartsen et~al., {\em Astrophys. J.} {\bf
  765} (2013) 55.

\bibitem{Aartsen:2018mxl}
{\bf IceCube} Collaboration, M.~G. Aartsen et~al., {\em Eur. Phys. J.} {\bf
  C78} (2018) 831.

\bibitem{Gaisser:2016obt}
T.~K. Gaisser, {\em J. Phys. Conf. Ser.} {\bf 718} (2016) 052014.

\bibitem{IceTopVeto_ICRC2019}
{\bf IceCube} Collaboration,  \pos{PoS(ICRC2019)445} (these proceedings).

\bibitem{Bai:2007zzm}
{\bf IceCube} Collaboration, T.~Gaisser et~al., {\it {IceTop/IceCube
  coincidences}},  in {\em {Proc. of the 30th ICRC, Merida, Mexico}}, vol.~5,
  pp.~1209--1212, 2007.

\end{thebibliography}\endgroup
\noindent
\footnotesize{The preparations for the surface enhancement are support by several agencies, such as the National Science Foundation (NSF) and the Helmholtz Association. This project has received funding from the European Research Council (ERC) under the European Union's Horizon 2020 research and innovation programme (grant agreement No 802729).}
%

\end{document}